\documentclass[lettersize,journal]{IEEEtran}
\usepackage[noadjust]{cite}
\usepackage[english]{babel}
\usepackage{blindtext}
\usepackage{booktabs} % For formal tables
\usepackage{epsfig,endnotes, enumerate}
\usepackage{subfigure}
\usepackage{graphicx}
\usepackage{url}
\usepackage{color}
\usepackage{balance}
\usepackage{amssymb,amsmath, listings, verbatim, float}
\usepackage[linesnumbered, ruled, vlined]{algorithm2e}
\usepackage{mathtools, comment}
\usepackage{multirow}
\usepackage{etoolbox}

\definecolor{blue}{rgb}{0.0,0.0,1}

\definecolor{red}{rgb}{1,0.0,0.0}

\definecolor{gray}{rgb}{0.3,0.3,0.3}
\newcommand{\code}[1]{\textcolor{gray}{\texttt{#1}}}

\begin{document}

\title{Transparent and Tamper-Proof Event Ordering in the Internet of Things Platforms}

\author{Mahbubur Rahman and Abusayeed Saifullah
% <-this % stops a space
\thanks{Mahbubur Rahman is with the City University of New York, and Abusayeed Saifullah is with Iowa State University.}}

%\markboth{IEEE Internet of Things Journal}%
{}

%\IEEEpubid{0000--0000/00\$00.00~\copyright~2022 IEEE}

\maketitle

\begin{abstract}
Today, the audit and diagnosis of the causal relationships between the events in a trigger-action-based event chain (e.g., why is a light turned on in a smart home?) in the Internet of Things (IoT) platforms are untrustworthy and unreliable. The current IoT platforms lack techniques for transparent and tamper-proof ordering of events due to their device-centric logging mechanism.
In this paper, we develop a framework that facilitates tamper-proof transparency and event order in an IoT platform by proposing a Blockchain protocol and adopting the vector clock system, both tailored for the resource-constrained  heterogeneous IoT devices, respectively.
To cope with the unsuited storage (e.g., ledger) and computing power (e.g., proof of work puzzle) requirements of the Blockchain in the commercial off-the-shelf IoT devices, we propose a partial consistent cut protocol and engineer a modular arithmetic-based lightweight proof of work puzzle, respectively.
To the best of our knowledge, this is the first Blockchain designed for resource-constrained heterogeneous IoT platforms. Our event ordering protocol based on the vector clock system is also novel for the IoT platforms. We implement our framework using an IoT gateway and 30 IoT devices. We experiment with 10 concurrent trigger-action-based event chains while each chain involves 20 devices, and each device participates in 5 different chains. The results show that our framework may order these events in 2.5 seconds while consuming only 140 mJ of energy per device. The results hence demonstrate the proposed platform as a practical choice for many IoT applications such as smart home, traffic monitoring, and crime investigation. 
\end{abstract}


\section{Introduction}\label{sec:intro}

%talk about the current IoT frameworks
%Internet of Things (IoT) applications are greatly influencing every aspect of our lifestyle, from home to a hospital bed. As of today, we have numerous smart home IoT platforms (e.g., Google Android Thing~\cite{google}, Samsung SmartThings~\cite{samsung}, Apple HomeKit~\cite{apple}, Iris~\cite{iris}) that help automate our home appliances. Similarly, smart health initiatives provide per patient monitoring which is more effective and adaptive~\cite{smarthealth}. Typically, such IoT platforms provide an application software for the management of the smart devices. Additionally, they expose a programming framework for third-party applications, enabling advanced device automation~\cite{samsung}. As such, smart devices and applications interconnect through the Internet or gateway and chain together to perform diverse range of activities.

Internet of Things (IoT) applications are greatly influencing every aspect of our lifestyle, including our activities at home, safety at public places and roads, and care in a hospital bed. 
As of today, there are numerous IoT platforms to automate our home appliances~\cite{logging1}, monitoring systems to automate traffic flows~\cite{cucchiara2000image, sichitiu2008inter, semertzidis2010video}, network deployments to ensure public safety~\cite{doumi2013lte, butun2016cloud}, and smart health systems for patient monitoring~\cite{boateng2019experience, park2017glasses, jia2017monitoring}. Typically, such IoT platforms allow smart sensors and/or applications to interconnect through the Internet or gateway and {\em chain together} to perform diverse activities. 
They also provide programming frameworks to enable advanced automation through chaining of multiple third-party applications.

Despite their configurability, many IoT platforms lack transparent and tamper-proof detection of {\em causal dependencies} between the sensors, especially during emergency/audit. For example, a traffic monitoring system may not provide a transparent scenario of an accident that involves one or multiple road intersections; a compromised public safety monitoring system may fail to order the events needed for a crime investigation; a smart health platform may not identify the root causes of a monitored patient going into critical condition; and a smart home platform may not conclude if a porch light is turned on because a motion sensor  has detected motion or the front door has been unlocked. The reasons behind these scenarios include {\em device-centric} logging mechanism, {\em vulnerability}, and {\em heterogeneity} of the IoT devices.

Although the gateway gets a centralized view of the whole platform by congregating logs from the devices, it is unable to construct accurate causal dependencies between the IoT devices/sensors due to the lack of synchronization between them~\cite{cheney2009provenance}. For example, consider the following high-level device logs provided by an Iris security system gateway: {\em ``motion detected by camera at 11:13 AM"}, {\em "front door unlocked at 11:13 AM"}, {\em ``porch light turned on at 11:14 AM"}~\cite{logging1}. This gateway thus cannot provide a causal dependency between these light, camera, and door sensors. As reported in~\cite{safetyissue}, ZigBee vulnerability lets hackers use hue bulbs to hijack any smart home, thus introducing trust issues as well. 
%Similarly, accident or crime investigation based on the sensor logs in any traffic or public safety monitoring system may not be accurate due to the device vulnerabilities.  
In general, the lack of a uniform ontology between the heterogeneous devices and uncertain temporal behavior in these systems make it extremely difficult to derive the causal dependencies.

%Talk about what are we trying to do 
%In this paper, we propose an IoT framework -- called {\em Transparent IoT (T-IoT)} -- where the causal relationships, i.e., the data provenance between heterogeneous IoT devices become transparent to the users. {\em Data provenance} is a holistic tracking of the causal relationships between a sequence of activities or events within a computing system. Adopting our T-IoT framework, the Iris platform in the previous example would thus be able to conclude that the light was turned on at 11:14 AM because of either the motion detection event at 11:13 AM or the door unlock event at 11:13 AM. While such transparency has been provided in many other domains including cryptocurrency (e.g., Bitcoin maintains provenance of its transactions~\cite{nakamoto2008bitcoin}) and retail corporation (e.g., Walmart maintains provenance of its pharmaceuticals~\cite{walmart}) by adopting different variants of the Blockchain protocol~\cite{haber1990time}, those techniques rely on both large memory and high computation power and hence cannot be applied to resource-constrained IoT devices.

In this paper, we propose an IoT framework called {\em Transparent IoT (T-IoT)}, where the causal relationships (i.e., data provenance) between heterogeneous IoT devices become transparent and tamper-proof. 
%T-IoT can benefit a number of IoT applications including smart home and traffic and public safety monitoring. 
Formally, {\em data provenance} is a holistic tracking of the causal relationships between a sequence of activities within a computing system. To design the core of T-IoT, we take motivation from the existing transparent and tamper-proof systems such as cryptocurrency (e.g., Bitcoin maintains provenance of its transactions~\cite{nakamoto2008bitcoin}) and retail corporations (e.g., Walmart maintains provenance of its pharmaceuticals and produce for safety and tracking~\cite{walmart}). While these platforms can afford resource-hungry 
%(e.g., large storage, high computing power, and high bandwidth)
{\em Blockchain} protocols to ensure tamper-proof transparency, it is not well-suited for the resource-constrained IoT devices. In T-IoT, we thus design a Blockchain protocol tailored for the resource-constrained IoT devices and enable the tamper-proof transparency of their event provenance.

Enabling Blockchain over resource-constrained IoT devices raises a number of practical challenges. In Blockchain (e.g., Bitcoin), each participating entity (e.g., miners) maintains the entire copy of a continually growing distributed ledger of transactions via a Byzantine consensus protocol -- called the {\em Nakamoto consensus} over a peer-to-peer (P2P) network to provide transparency of the transactions. The correctness of such consensus depends on a {\em computationally expensive} proof of work (PoW) protocol~\cite{nakamoto2008bitcoin}. The PoW protocol and the ledger prevents the {\em double spending} problem (spending the same coin more than once by tampering the ledger) in the Bitcoin ecosystem. In T-IoT, {\em events} are analogous to the Bitcoin {\em transactions} and the {\em double spending} refers to an {\em inconsistency} in its data provenance, e.g., the root cause of an event referring to several IoT nodes. In the same spirit of Bitcoin, T-IoT maintains a ledger of its events and employs a PoW protocol. 
Specifically, the design of T-IoT addresses the following key practical challenges.

% \begin{enumerate}
% 	\item Commercial off-the-shelf (COTS) IoT devices provide only a few hundred KB of flash memory (e.g., 128 KB in TI CC1310), which is again shared between the system and application programs. It is thus impractical for the IoT devices to participate in a Blockchain protocol where the ledger grows continually (currently, 270+ GB in Bitcoin~\cite{bitcoinsize}).

% 	\item COTS devices typically perform ultra low-power operation and need to have a battery-life of several years, thereby making them naturally unsuited for the PoW protocol that has to rely on high computing power, time, and energy budget~\cite{bitcoinenergy}. 

% 	\item Depending on their functionality, different IoT devices are equipped with different wireless communication protocols (e.g., Wi-Fi, BLE, ZigBee, or LoRa). Such heterogeneity makes it difficult to develop a P2P protocol to enable the Nakamoto consensus protocol.

% 	\item The lack of a common ontology between the devices from different vendors makes it difficult to synchronize them. As a result, deriving any cause-effect relationship between the events in an IoT platform becomes extremely difficult.
% \end{enumerate}

(1) Commercial off-the-shelf (COTS) IoT devices provide only a few hundred KB of flash memory (e.g., 128 KB in TI CC1310), which is shared between the system and application programs. It is thus impractical for the IoT devices to participate in a Blockchain protocol where the ledger grows continually (currently, 375+ GB in Bitcoin~\cite{bitcoinsize}).
(2) COTS devices typically perform ultra low-power operation and need to have a battery-life of several years, thereby making them naturally unsuited for the PoW protocol that has to rely on high computing power, time, and energy budget~\cite{bitcoinenergy}. 
(3) Depending on their functionality, different IoT devices are equipped with different wireless communication protocols (e.g., Wi-Fi, BLE, ZigBee, or LoRa). Such heterogeneity makes it difficult to develop a P2P protocol to enable the Nakamoto consensus protocol.
(4) The lack of a common ontology between the devices from different vendors makes it difficult to synchronize them. As a result, deriving any cause-effect relationship between the events in an IoT platform becomes extremely difficult.
In this paper, we address the above challenges and enable transparent and tamper-proof event ordering in the IoT platforms. The key novel contributions of this paper are as follows.
%The key to our framework is a Blockchain protocol, which, to the best of our knowledge, is the first Blockchain protocol tailored for the resource-constrained heterogeneous IoT devices. Specifically, we make the following contributions.

%\vspace{-0.05in}
\begin{itemize}
	\item We enable Blockchain in T-IoT by allowing each node to save only a {\em portion of the ledger} that relates to the most recent events in the platform.
	%s, thus handling the storage limitation. 
	As the ledger grows, a node replaces its portion over time. To do this, we propose a {\em partial consistent cut}-based replacement policy that finds the dependencies between multiple events (based on the cut size) that occurred in the platform. We also propose a {\em modular arithmetic}-based lightweight PoW protocol that is computationally fast for the IoT nodes.

	\item We enable the ordering of events by {\em logically} synchronizing the nodes. For this, we extend Lamport's logical clock~\cite{lamport1978ordering} to vector clocks, tailored for the IoT platforms. We then propose a backtracking-based algorithm to create the data provenance in T-IoT. 
	Additionally, we enable a gateway-assisted P2P communication in T-IoT. 
	%To the best of our knowledge, T-IoT is the first to logically synchronize heterogeneous IoT nodes.

	\item We evaluate the performance of T-IoT indoor. The gateway is implemented on GNU Radio using USRP (Universal Software Radio Peripheral) devices to support various communication protocols. We deploy 30 nodes (19 TI CC1310s with IEEE 802.15.4g, 3 TI CC1350s with BLE, and 8 Dragino LoRa nodes) in our testbed. We then activate 10 trigger-action-based event chains. Each chain involves up to 20 nodes, and each node may participate in 5 different chains. Our results show that when 10 chains execute concurrently, the ordering of their events may be done in 2.5 seconds at the cost of $\approx$140 mJ of energy per node, thus demonstrating the feasibility of T-IoT.
\end{itemize}

In the rest of the paper, Sections~\ref{sec:background} and~~\ref{sec:overveiw} overview our system model and design, respectively. Sections~\ref{sec:blockchain},~\ref{sec:ordering}, and~\ref{sec:provenance} detail the T-IoT Blockchain, event ordering, and provenance creation protocols, respectively. 
Sections~\ref{sec:imple} and~\ref{sec:experiments} provides the implementation details and the performance evaluation of different protocols of our framework. Section~\ref{sec:related} overviews the related work. Finally, Section~\ref{sec:conclusion} concludes our paper.

\section{System Model and Background}\label{sec:background}
In this section, we discuss our system model and provide background knowledge on Blockchain and data provenance.

\begin{figure}[!htbp]
\centering %\vspace{-0.1in}
\includegraphics[width=0.36\textwidth]{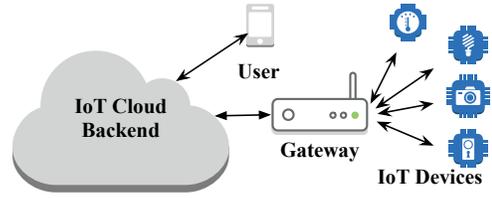}%\vspace{-0.1in}
\caption{Network model of T-IoT.}%\vspace{-0.1in}
\label{fig:iotplatform} %\vspace{-0.1in}
\end{figure}
\subsection{System Model of T-IoT}\label{sec:network_model}
\noindent{\bf Network Model.}
%Briefly talk about the network model and why this kind of network model
Figure~\ref{fig:iotplatform} shows the network model of T-IoT, which represents the existing IoT platforms. It consists of a variety of IoT devices including its users, a gateway or hub, and a Cloud backend. 
The devices are heterogeneous and have limited power (e.g., battery-powered) and storage capability to log their activities. Each device is equipped with one radio front-end (e.g., Wi-Fi, BLE, Zigbee, etc.) to send/receive commands to/from the gateway.
%The gateway is wall-powered and manages the IoT devices. The gateway 
The gateway is computationally powerful, wall-powered, and is connected to the Internet/Cloud. It manages each device through a {\em device abstraction layer}. It is equipped with heterogeneous wireless transceiver radios that allow it to communicate with the heterogeneous devices. Also, it connects to the Cloud using Wi-Fi or Ethernet.
The users can host different applications (e.g., smart home) using the Cloud backend. The Cloud also acts as a data storage for the applications. 
Such an architecture enables an automatic management of a target application where users can remotely enable, monitor, and control various activities that involve trigger-action based chaining of numerous devices.

\noindent{\bf Assumptions and Facts in T-IoT.}
In this paper, we consider that the T-IoT gateway and the Cloud backend are the trusted entities. Securing the gateway and the Cloud backend is out of the scope of this paper. Instead, we design novel protocols to ensure tamper-proof and transparent ordering of the events of the resource-constrained IoT devices (e.g., sensors) that are {\em more vulnerable and major entry point} to the adversaries~\cite{safetyissue, hijack2}. We also assume that, at any given point of time, more than half of the IoT devices will function properly (e.g., no hardware failure/compromised) in T-IoT to facilitate the correct ordering and tamper-proof transparency of the events.
Despite having a centralized view and sufficient storage and computing power, the gateway may not construct a causal ordering of the events in T-IoT. The reason is that the gateway entirely depends on the IoT devices' logs to learn at what exact time the events occur. Also, even within a single trigger-action-based event chain, the lack of synchronization between the IoT devices may alter the causal ordering in the chain. The presence of multiple chains with one or more common IoT devices makes this scenario more complicated. Despite these issues, even if the gateway can order events on its own (for argument's sake), T-IoT will lose the tamper-proof property.

From the security perspective, the major breakpoint for both the gateway and Cloud is related to the authentication (e.g., passwords of less complexity) and transport/network layer encryptions, which is common to the existing wireless/wired systems. The resource-constrained IoT devices, on the other hand, are vulnerable to a variety of issues that are related to the device proximity (e.g., physically compromising motion sensors and security cameras), hardware (e.g., connecting to JTAG UART/I2C/SPI of the system to generate false alarms), and protocol stack (e.g., using man-in-the-middle and replay attacks to tamper device logs)~\cite{hijack3}. These vulnerabilities inspire the need for a tamper-proof protocol (e.g., Blockchain) in the resource-constrained IoT devices even if the gateway and Cloud are trusted.
%We thus limit our focus to the tamper-proof transparency of the events in T-IoT.}

\subsection{Background Knowledge}\label{sec:provdm}
%In the following, we briefly discuss the Bitcoin Blockchain protocol and W3C PROV-DM data provenance model~\cite{provmodel}. 
%In the following, we briefly discuss the Bitcoin Blockchain protocol which we tailor for the resource-constrained IoT devices. Additionally, we discuss the W3C PROV-DM data provenance model~\cite{provmodel} that we use to formally represent the transparent event ordering in T-IoT. 

\noindent{\bf Bitcoin Blockchain.}
%talk about bitcoin and blockchain structure
Bitcoin is a cryptocurrency and used by the interested parties to complete financial transactions without a central administrator (e.g., Banks). To facilitate transactions, Bitcoin ecosystem creates a P2P network of its parties (say, nodes). Transactions are verified by the network nodes and then recorded in a public distributed ledger -- called a {\em Blockchain}. The Blockchain thus holds the records of the Bitcoin transactions in the form of a growing list of records called {\em blocks} that are linked using cryptography. Each block in the Blockchain contains a cryptographic hash of its parent block, control information (e.g., timestamp, transaction hash, etc.), and transaction data in the form of a Merkle tree to facilitate creation of their hash value~\cite{nakamoto2008bitcoin}. 
Only a group of special nodes (called {\em miners}) in the network can add new blocks (called {\em block validation/mining}) to the existing chain. To add a new block, a miner solves a computationally expensive puzzle known as the PoW and other miners have to agree that the solution is acceptable via a Byzantine consensus protocol known as the {\em Nakamoto consensus}~\cite{nakamoto2008bitcoin}. All the miners have to maintain a complete copy of the already validated chain to participate in and provide correctness of this process. Once a block is mined, each miner adds that block to its chain without requiring any central oversight. The Blockchain, by design, thus becomes resistant to modification of the transaction data and prevents the double spending problem at the cost of extensive computation and storage in the miner, as long as majority of the miners are honest.%Nonetheless, because of its key features including the distributed ledger, P2P network, and the PoW protocol, the Blockchain protocol appears as a promising  technology for building the data provenance in the existing IoT platforms. 
\begin{figure}[!htbp]
\centering %\vspace{-0.1in}
\includegraphics[width=0.35\textwidth]{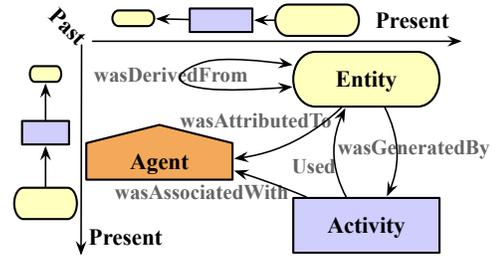} %\vspace{-0.1in}
\caption{PROV-DM provenance model.} %\vspace{-0.1in}
\label{fig:prov} 
\end{figure}

\noindent{\bf Data Provenance.}
%define data provenance and why it is useful in our framework
In T-IoT, we enable a tamper-proof transparent ordering of the IoT events. Such ordering is useful to the users when its {\em knowledge representation} is pervasive and easily comprehensible. We thus present the order of events in the form of a data provenance. Data provenance systematically describes the history of the actions taken on an object (e.g., data, event, entity, etc.) from its creation up to the present. Such knowledge presentation can answer many historical questions about an object, including {\em ``what entity triggered event $e_i$?"} and {\em ``how is event $e_j$ derived from event $e_i$?"}, which is useful in system diagnosis/audit~\cite{chen2016good, bates2015trustworthy, ma2016protracer}.

% \begin{figure}[t]
% \centering %\vspace{-0.1in}
% \includegraphics[width=0.47\textwidth]{figs/overview.eps} \vspace{-0.1in}
% \caption{Block diagram of the three protocols of T-IoT (dashed-arrows represent functional dependencies).} \vspace{-0.25in}
% \label{fig:overview}
% \end{figure}
% talk about the data provenance model used in this paper. 
We use the W3C PROV-DM~\cite{provmodel, nwafor2017towards} model to represent the event order in T-IoT. It represents provenance in the form of a directed acyclic graph (DAG) that consists of entity, activity, and agent nodes. An {\em entity} is a data object and may refer to many other entities. An {\em activity} is a process and defines how entities come into existence.
An {\em agent} bears responsibility for activities and entities. In short, such structure can describe a relationship in the following form: {\em ``the agent was responsible for the activity which generated the entity"}. The edges in PROV-DM DAG encode a variety of dependencies between the nodes, as shown in Figure~\ref{fig:prov}, where the timeline follows past to present from left to right and top to bottom. 
In general, using the PROV-DM in T-IoT, we may enable the T-IoT users to learn (without having to deal with the low-level and obtuse logs from the devices/gateway) which entity \code{wasDerivedFrom} which entity, which entity \code{wasGeneratedBy} which activity, which activity \code{used} which entity, which activity \code{WasAssociatedWith} which agent, and which entity \code{WasAttributedTo} to which agent.

\section{T-IoT Framework Overview}\label{sec:overveiw}
\begin{figure}[!htbp]
\centering %\vspace{-0.1in}
\includegraphics[width=0.49\textwidth]{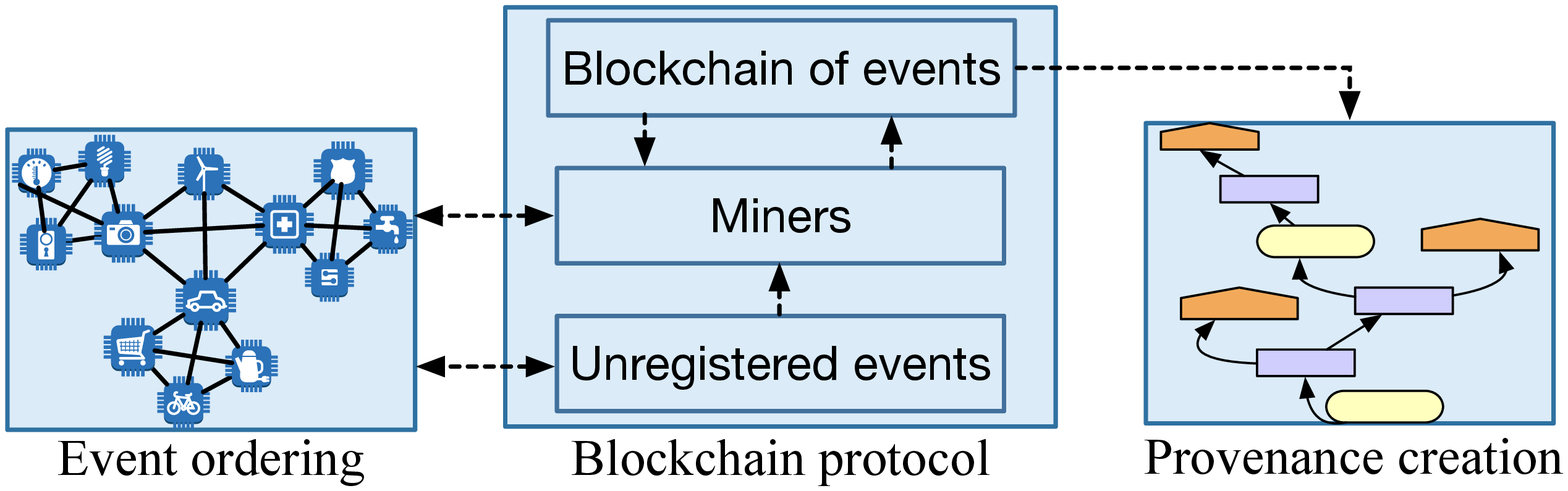} %\vspace{-0.1in}
\caption{Block diagram of the three protocols of T-IoT (dashed-arrows represent functional dependencies).} %\vspace{-0.25in}
\label{fig:overview}
\end{figure}

In this section, we briefly overview the T-IoT framework design, which is depicted in Figure~\ref{fig:overview}.
%T-IoT consists of three major protocols: Blockchain, event ordering, and provenance creation. 
%In the following, we provide a brief description of these protocols and discuss their dependencies.

\noindent{\bf Blockchain protocol.} In T-IoT, we create a distributed ledger of the events that are generated by the IoT nodes. An event is said {\em unregistered} as long as it is not added to the ledger. Several IoT nodes act as miners to add all these events in the ledger (as blocks of events). This process is called {\em event registration} (i.e., mining). 
All the events within an event chain are initially unregistered. In time, all of them become registered (added to the ledger).
The gateway plays a vital role in the trigger-action-based event chain by managing (e.g., sending event commands) the IoT nodes. 
It allows an action event only if all the triggering events are already registered.
Due to numerous trigger-action-based chains in the system, an IoT node may be involved in multiple chains, and thus maintains a list of the registered events (i.e., the ledger). To do so, it saves only a portion of the ledger (which is updated over time) because of its storage limitation. The gateway has sufficient storage (since its connected to the Cloud) and saves the entire ledger. 

\noindent{\bf Event Ordering Protocol.}
Our event ordering protocol runs in parallel to the event registration process. All the unregistered events are listed in one/more blocks by one or more miners in the order they are generated. To achieve such ordering, we logically synchronize the IoT nodes by extending Lamport's logical clock system. In this process, each miner maintains a vector of event's count to track the number of event requests made by other miners that have at least one common event with it. In this way, a miner waits to group an event in a block until the other miners confirm (by sending messages through the gateway) that there are no unknown preceding events.

\noindent{\bf Provenance Creation.}
The gateway creates the data provenance in the form of a DAG. For each trigger-action-based event chain, it identifies the start event (root cause) and the end events (final effects) of that chain and builds a provenance graph while conforming to the PROV-DM model. Since the blocks in the ledger contain events in an orderly fashion, the gateway thus starts from the most recent block to find the effects and then backtracks to as many blocks as needed.
%to find the root cause.

\section{T-IoT Blockchain Protocol}\label{sec:blockchain}
In this section, we detail the T-IoT Blockchain protocol that provides tamper-proof transparency in the IoT platforms.

\subsection{Blockchain Primers}\label{sec:b_primers}
%We first discuss about several primers of the T-IoT blockchain such as the virtual currency, distributed ledger, and message passing between the IoT nodes. 

\noindent{\bf Virtual Currency.}
In T-IoT, we use {\em virtual coins} (or, simply {\em coins}) in transactions (e.g., event requests) between the nodes and the gateway. Coins are necessary to limit the number of event requests from the nodes to the gateway. A node spends a coin (i.e., pays to the gateway) to request an event. 
In other words, a node pays the gateway with a coin to initiate an event.
There is thus no notion of fractional coin transfer in T-IoT. 
A node is assigned a fixed integer number of coins when it joins the network. Specifically, a node gets $N$ coins if it may generate $N$ distinct events.
For example, a door sensor in a smart home gets 2 coins if it can activate (via the gateway) two different light bulbs. The gateway restores a spent coin if the associated event is validated (detailed in Section~\ref{sec:tx_details}).
At any given point of time, the total number of coins in T-IoT is thus fixed depending on the number of events.

\noindent{\bf Distributed Ledger.}
In T-IoT, we maintain a distributed ledger of the events generated by the nodes, which grows in size over time. The gateway saves the entire ledger while each node saves a portion of it. Such a design decision is made due to the following two reasons. (1) The nodes are memory-constrained, and it is impractical for them to maintain an ever-growing ledger. (2) The memory or storage at the gateway is not a big concern since it is connected to the Cloud. In its partial ledger, each node saves the most recent events, specifically the events that are generated by its associated nodes.
For example, a light sensor saves the recent events of a motion sensor and a door sensor if these sensors can generate an event in it.
A node learns about the associated nodes when it joins the network (e.g., during its installation by a user/technician through manual/Cloud configuration).

\noindent{\bf Message-Passing.}
In T-IoT, we enable P2P communications between any two nodes via the gateway. Since the gateway is equipped with heterogeneous receiver and transmitter radios, the communication between two nodes with different protocols (e.g., between ZigBee and BLE) is thus possible.
Specifically, a message delivery between a sender and a receiver (or a set of receivers) happens in the following two steps. 
(1) The sender sends the message to the gateway. (2) The gateway then broadcasts the message in the network. 
Depending on the IoT platform, the nodes may adopt {\em Low Power Listening}~\cite{lpl} or on-board sensor-triggered wake-up policy~\cite{onboard1} to listen to the broadcast messages with ultra-low energy consumption. This may also help the T-IoT framework to detect malfunctions/vulnerabilities under very low-traffic or low device-activities.

\subsection{Transaction Details}\label{sec:tx_details}
%how coins are spent and restored ()
\noindent{\bf Transaction.}
An event request by a node to the gateway is a transaction in T-IoT. A node pays the gateway with a coin for each event request. 
T-IoT does not incur any fee for transactions, as it bears no meaning. Similarly, 
there is no incentive (reward coin) for the nodes that validate such transactions. Adding an event to the ledger means that the gateway has allowed that event to execute.  A unique {\em event\_handler} or a number represents each event in T-IoT. Adding an event to the ledger thus refers to adding the associated number.
In each transaction, a node incorporates its event request and all the validated events of the trigger-action-based event chain it is involved in. 
For simplicity, the {\em event\_handler} of each event in T-IoT represents the coin for itself. Paying the gateway for an event thus refers to the inclusion of the {\em event\_handler} in the event request.
The gateway tracks the validity of the coin by associating one additional {\em bit} of information (say, {\em coin\_bit}) with each {\em event\_handler} where 0 means the payment is valid (i.e., event request is valid) and 1 means the payment is invalid.
For a valid request, it sets the {\em coin\_bit} and then broadcasts the transaction in the network for validation. For a successful event validation (discussed in Section~\ref{sec:pow}), the gateway resets the corresponding {\em coin\_bit} so that the requesting node may be able to register the same event in the future. For an invalid payment (thus a replay attack), the gateway rejects the event request by checking if the associated {\em coin\_bit} is set.

\begin{figure}[!htbp]
\centering %\vspace{-0.15in}
\includegraphics[width=0.4\textwidth]{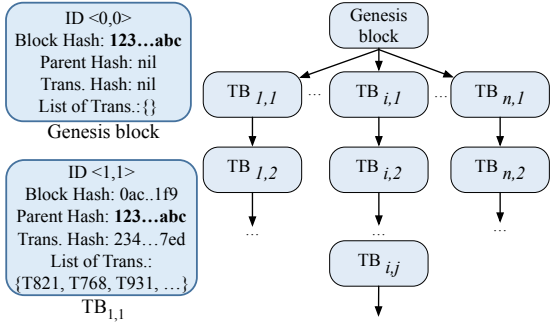} %\vspace{-0.15in}
\caption{A generic view of transaction block (TB) and ledger in T-IoT.} %\vspace{-0.07in}
\label{fig:ledger} %\vspace{-0.05in}
\end{figure}
% a block detail
\noindent{\bf Transaction Block.}
Multiple transactions are grouped together to create a block based on two criteria. (1) The events belonging to the same chain are included in the same block (a node knows its scope). (2) If a node belongs to multiple chains, events in these chains that happen concurrently at any given point of time are also included in the same block.
Thus, there may be several chains of blocks rooted at the {\em genesis block}, as shown in Figure~\ref{fig:ledger}, depending on how the IoT devices are chained together to create the trigger-action-based chains. A genesis block in any Blockchain protocol refers to the very first block in the Blockchain and serves as an entry point of search through the Blockchain (similar to the head pointer of a linked list). Having multiple chains of blocks reduces the search space at the gateway, which makes it highly scalable. 
In a transaction block, several other control information such as ID, parent hash, transaction's hash are also added (Figure~\ref{fig:ledger}) to maintain the integrity and structure of the ledger.
%Note that such a Blockchain structure helps the memory-constrained nodes save a complete or partial chain of blocks of its related events.

\subsection{The PoW Protocol}\label{sec:pow} 

The T-IoT PoW protocol includes identifying the miners, dealing with the resource
%(e.g., storage and computing power) 
limitations, and block validation.

\noindent{\bf Miner Identification.}
%In Bitcoin system, it is necessary that miners have sufficient resources (e.g., computational power and storage) to validate a transaction block to complete the proof of work. As a result, Bitcoin grantees trust and transparency to its users. T-IoT, as a transparent framework, also requires a proof of work protocol.
Any Blockchain protocol guarantees tamper-proof transparency using a PoW protocol (or smart contract) that requires miners~\cite{nakamoto2008bitcoin, choudhury2018enforcing}.
In T-IoT, a node with one of the following criteria participates as a miner to validate a transaction block of unregistered events. (1) At least one event in the transaction block is common to the events in the trigger-action-based event chains it belongs to. (2) T-IoT has made an earlier reservation with a node to act as a miner.
While the first criterion is intuitive, the second criterion is added in our framework to support isolated events that do not belong to any trigger-action-based event chain.

\begin{figure*}[!htbp]
\centering %\vspace{-0.2in}
\includegraphics[width=0.6\textwidth]{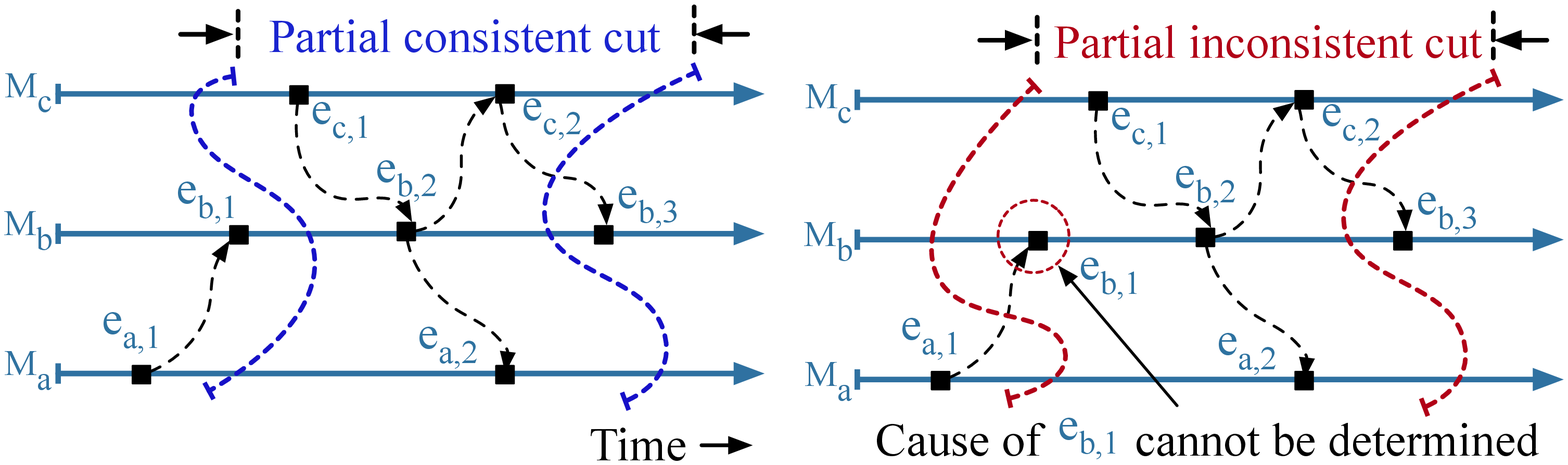} %\vspace{-0.3in}
\caption{Left figure shows one of several partial consistent cuts (two dotted negative-ended curved lines) of the events (denoted by black squares and labeled accordingly) of 3 miners M$_a$, M$_b$, and M$_c$.} %\vspace{-0.1in}
\label{fig:cut} 
\end{figure*}
\noindent{\bf Dealing with the Storage Requirements.}
%talk about consistent cuts
%talk about block replacement policies
%In T-IoT, it is very challenging for the resource-constrained (e.g., limited storage capacity and computation capability) miners to participate in the PoW. In the following, we discuss how we overcome these limitations.
As discussed in Sections~\ref{sec:b_primers} and~\ref{sec:tx_details}, a miner may not be able to store the entire ledger that grows in size over time. Consequently, depending on its storage capacity, a miner saves several of the recently validated blocks of its related events. In general, the goal is to efficiently use a miner's limited storage capacity so  that it can effectively participate in the T-IoT Blockchain protocol. 
%The cut in the right figure is not a partial consistent cut.
While saving a block into its storage, a miner maintains a partial consistent cut of the chain of blocks of its related events. Figure~\ref{fig:ledger} shows an example of such a chain of blocks as $\{$TB$_{i, 1} \rightarrow$ TB$_{i, 2} \rightarrow \cdots \rightarrow$ TB$_{i, j} \rightarrow \cdots\}$. 

A {\em partial consistent cut} of a chain refers to the portion (i.e., the subsequence of blocks) of the chain, where each event in that portion is traceable to its triggering event, as shown in Figure~\ref{fig:cut}. In this figure, the chain has three miners M$_a$, M$_b$, and M$_c$, executing events $\{$e$_{a,1}$, e$_{a,2}\}$, $\{$e$_{b,1}$, e$_{b,2}$, e$_{b,3}\}$, and $\{$e$_{c,1}$, e$_{c,2}\}$, respectively, and one of the partial consistent cuts (as shown on the left side of this figure) includes events $\{$e$_{c,1}$, e$_{c,2}$, e$_{b,2}$, e$_{a,2}\}$ from a few blocks in the chain. Note that inclusion of event e$_{b,3}$ to this cut will also be another partial consistent cut. A miner thus saves a portion of the ledger using this technique and the cut size will be determined by its storage size. If a miner can save $M$ blocks and each block contains on average $E$ events, then the worst-case time complexity for finding a partial consistent cut is $O(ME)$. The partial cut shown on the right side of Figure~\ref{fig:cut} is an {\em inconsistent} cut since event e$_{b,1}$ cannot be traced back to its triggering event. 

In time, a miner replaces an older block with a newer block while maintaining a partial consistent cut. A miner may also save the most recent blocks that are not included in its partial consistent cut, as long as it has storage capacity. For an event validation, a miner may also request the gateway for the missing block/s (a block fits in the payload of one packet, as discussed in Section~\ref{sec:setupchain}), which is analogous to the notion of the {\em cache} and {\em main memory} in the CPUs.
%As we shall see in , a block fits in the payload of one packet transmission. 
It saves the incoming block/s while maintaining a partial consistent cut or using a cache replacement policy such as the {\em least recently used} (LRU). The LRU technique may also be used if no partial consistent cut exists. In contrast to the other cache replacement policies, the LRU policy is beneficial for a node since the recent blocks are most suited for new event validation.

%The node may save the received blocks (by replacing older blocks), while maintaining a partial consistent cut or using a cache replacement policy such as the {\em least recently used} (LRU) technique. The LRU technique may also be used if no partial consistent cut exists for a miner.

\noindent{\bf Puzzles and Dealing with the Computation Requirements.} 
Our goal here is to efficiently use the limited computing power of the miners so that they can solve the PoW puzzles quickly and energy efficiently. In T-IoT, the gateway generates one puzzle per event. The rationale behind a puzzle per event is twofold. (1) No miner can dominate the T-IoT PoW protocol. (2) Each event becomes tamper-proof so that a compromised miner cannot falsify it with a group of events. In the following, we explain the strategy to generate a puzzle for event $e$. 

The gateway chooses a large prime number $P$ and calculates the {\em primitive roots} of $P$. A primitive root of a prime $P$ is an integer $r$ which is relative prime with $P$ and $r~(\text{mod}~P)$ has a {\em multiplicative order} $(P -1)$~\cite{ribenboim2012new}. The number of primitive roots of $P$ is exactly $\phi (\phi(P))$, where $\phi$ is the {\em Euler phi function}~\cite{lehmer1932euler}. For prime $P$, $\phi(P) = P-1$. For any other positive integer $N$, $\phi(N) = N \Pi_{p|N}(1 - \frac{1}{p})$, where $p$ is a prime factor of $N$.
After calculating the primitive roots for event $e$, the gateway assigns a {\em unique} tuple $\langle P, r_i \rangle$ to each of the miners associated with event $e$, where $r_i \in \{ r_1, r_2, r_3, \cdots, r_{\phi (\phi(P))} \}$ is the $i$-the primitive root of prime $P$. Note that a miner which is associated with $m$ distinct events will thus get $m$ tuples $\langle P_j, r_{jk} \rangle$, where prime $P_j$ is distinct for $m$ distinct events and $r_{jk} \in \{ r_{j1}, r_{j2}, r_{j3}, \cdots, r_{j\phi (\phi(P))} \}$ is the $k$-th primitive root of $P_j$.
During the block creation (containing event $e$), a miner that is assigned tuple $\langle P, r_i \rangle$ tries to solve the puzzle: $(r_i)^K~\text{mod}~P = r_i$ for $K$, where $K > P * rand(1, r_i)$. The $rand(1, r_i)$ function generates a random value
within the range between between 1 and  $r_i$, which brings unpredictability in the system and allows different miners to solve a puzzle for the same event occurring at different times in the system.
%For $n$ miners of event $e$ with chosen prime $P$, the gateway calculates different primitive roots in $O(n.\log \phi(P).\log P)$ time.

A miner, on the other hand, tries to a solve the puzzle for event $e$ computationally efficiently using {\em Fermat's Little Theorem}~\cite{burton1985history}. According to Fermat's Little Theorem, if $r$ is relative prime with $P$ and $P$ is a prime number, then $r^{P-1} \equiv 1~(\text{mod}~P)$.
During block validation (containing event $e$) process, the other miners associated with event $e$, each having a different tuple $\langle P, r_j \rangle$, assure the correctness of the solution by checking if $(r_j)^K~\text{mod}~P = r_j$ holds. Such a mathematical relationship is consistent between the primitive roots of a prime and also used in {\em Diffie--Hellman key exchange}~\cite{diffhill}.
We provide an execution-time estimation for solving such a puzzle with different primes and their primitive roots in Table~\ref{tab:proots} using three IoT devices: TI CC1310 and CC1350 (both have 32-bit Cortex-M3~\cite{cc1310}) and LoRa Hat on Raspberry Pi (RPi) 3 (64-bit Quad-Core 1.2GHz~\cite{raspberrypi}). For the 1000th prime, the value of $K$ is found within hundreds of ms using TI CC1310/CC1350 and only tens of ms using RPi. The T-IoT PoW puzzles are thus very efficient for IoT nodes. {While this PoW may be vulnerable to Shor's factor decomposition algorithm~\cite{shorsalgo}, it would require a compromised {\em resource-constrained} IoT device to have the {\em capabilities of a quantum computer} (which is impractical).}
\begin{table}\footnotesize
\centering
\begin{tabular}{|c|c|c|c|c|c|c|}
\hline
\multicolumn{2}{|c|}{\textit{\textbf{Prime}}}   & \multicolumn{2}{c|}{\textit{\textbf{Primitive Root}}} & \multirow{2}{*}{\textit{\textbf{K}}} & \multicolumn{2}{c|}{\textit{\textbf{Time for K (ms)}}} \\ \cline{1-4} \cline{6-7} 
\textit{\textbf{Nth}} & \textit{\textbf{Value}} & \textit{\textbf{\# of Roots}}    & \textit{\textbf{Chosen}}    &                                      & \textit{\textbf{TI}}  & \textit{\textbf{RPi}}  \\ \hline
100                   & 541                     & 144                     & 360                         & 1081                                 & 26.51                        & 1.49                    \\ \hline
200                   & 1223                    & 552                     & 926                         & 2445                                 & 71.38                        & 1.90                    \\ \hline
400                   & 2741                    & 1088                    & 2520                        & 5481                                 & 191.8                        & 9.69                    \\ \hline
600                   & 4409                    & 2016                    & 2921                        & 8817                                 & 331.7                        & 16.59                   \\ \hline
800                   & 6133                    & 1728                    & 5264                        & 12265                                & 507.3                        & 24.87                   \\ \hline
1000                  & 7991                    & 3816                    & 3926                        & 15837                                & 677.1                        & 32.62                   \\ \hline
\end{tabular}\vspace{0.05in}
\caption{Time to find K.}
\label{tab:proots}%\vspace{-0.35in}
\end{table}

\noindent{\bf Block Validation.}
This process adds blocks to the ledger. It starts when a miner (among many) claims (to the gateway) that it has created a block by solving all the related puzzles in the block. 
%Such a claim is broadcasted to the network via the gateway.
The gateway then broadcasts this block to the {\em associated} miners.
%Note that other miners also try to solve similar puzzles and create blocks. 
All these miners at this point stop creating their own blocks, check the correctness of the puzzles, and inform the gateway. 
%To this extent, if the gateway finds that no less than 51\% miners agree to the correctness of the puzzles, it will add the block to the ledger and broadcast this information in the network.
The gateway then adds the block to the ledger only if no less than 51\% of these miners agree to the correctness of the puzzles.
The overall process is refereed to as {\em Nakamoto consensus}. Once a block is added to the ledger, each miner resumes its block creation process and disregards events of the newly validated block. If multiple blocks are created by different miners at the same time, the gateway broadcasts the block with a greater number of events in it, where ties are broken randomly. During block validation of the isolated events (that do not belong to any event chain), there may be only one active miner. 
In this case, the gateway checks the correctness of the solved puzzles.

\section{T-IoT Event Ordering Protocol}\label{sec:ordering}
In this section, we discuss how the miners create a causal order of the events in T-IoT. 
%We first discuss why such an ordering cannot be achieved in the existing IoT platform. We then discuss our proposed ordering protocol.

%\noindent{\bf Why is Causal Ordering Difficult at the Nodes and Gateway?}
\subsection{Why is Causal Ordering Difficult at the Nodes and Gateway}
The Blockchain technology does not guarantee the causal ordering of the events at the nodes or gateway. {\em In general, canonical ordering of transactions (within a block) in the Blockchain is an active area of research}~\cite{daian2019flash, decker2014bitcoin, canonical}.
Due to the variable communication delays in event propagation and imperfect physical clocks in the IoT nodes, a particular event may arrive at different miners at different times. As a miner groups unregistered events in a block, it may not thus represent the global ordering of events. For example, depending on the arrival times of two events e$_1$ and e$_2$, two miners may group these events in the orders $($e$_1$, e$_2)$ and $($e$_2$, e$_1)$, respectively. Thus in the ledger, the order of these events will depend on whether the block from miner m$_1$ or m$_2$ is validated.
The jitter in event propagation originates from the gateway-assisted message passing protocol.
A lone gateway-based event ordering, if possible in a complicated event chain, will loose the temper-proof property of T-IoT.
Had we facilitated direct messaging between heterogeneous IoT nodes using the cross-technology communication~\cite{chen2019survey}, such jitters would still persist due to their conversion delays. A platform with a uniform ontology (e.g., only BLE) may also suffer due to imperfect physical clocks of the miners. Achieving {\em physical clock synchronization} may not solve this problem. Rather, it may limit the scalability. We thus focus on a {\em logical synchronization} in T-IoT by extending Lamport's logical clock~\cite{lamport1978ordering}, which is a practical choice for ordering events in a heterogeneous IoT platform.
%for the IoT platforms. Adopting the ordering protocol, each miner in T-IoT groups unregistered events in a block according to their global order. 
In the following, we first overview Lamport's logical clock.
%and then extend it to propose a vector clock-based distributed event ordering protocol.

%Prolem aries due to commuication delays of multicast, does not matter whether it comes via the gateway or in a p2p-multicast fashion directly between the nodes. In a validated block, a gateway can only see events, but not aware of the ordering at all.
%nodes may have actual event ordering inside a to-be validated block, thus the gateway can follow. Here we need the nodes to have the ordering solved in completely distributed way. 

%\noindent{\bf Lamport's Logical Clock.}
\subsection{Lamport's Logical Clock}
%talk about lamports logical clock and create a protocol similar to vector timestamp protocol, but it will be different. Talk about what events/transactions are incorporated within a block. May have to queue some events.
In 1978, mathematician and computer scientist Leslie Lamport showed that event ordering via synchronization between the nodes in a distributed system {\em need not} be based on the absolute time or physical clocks~\cite{lamport1978ordering, lamport2019time}. Even if two nodes do not interact, they should still be synchronized not necessarily because the lack of it will not be observable and thus may not cause problems, but rather it is related to the ordering of the events. 
Additionally, he argues that what suffices for the nodes to agree on is in what order the events occur (rather than the absolute time). In accordance, he defined a {\em ``happens-before"} relationship without referencing to the physical clocks. If $a$ and $b$ are two events, then ``$a$ happens-before $b$" means that all processes agree that first event $a$ occurs, then afterward, event $b$ occurs. Formally,
{\em happens-before} relation is defined as follows (denoted by ``$\rightarrow$").
%topsep=0pt, leftmargin=*
\begin{enumerate}[-]
	\item If $a$ and $b$ are two events in a process and event $a$ comes before event $b$, then $a \rightarrow b$.
	\item If $a$ is a message sending event in a process and $b$ is the receipt of that message by another process, then $a \rightarrow b$.
	\item If $a \rightarrow b$ and $b \rightarrow c$ then $a \rightarrow c$ for events $a$, $b$, and $c$. Two distinct events $a$ and $b$ are {\em concurrent} if $a \nrightarrow b$ and $b \nrightarrow a$.
\end{enumerate}

To facilitate this within a system, he introduced a {\em logical clock} that assigns a number to an event, where the number represents the time at which it occurs. Specifically, he defined a clock $\mathcal{C}_i$ for each process $p_i$ to be a function that assigns a number $\mathcal{C}_i \langle a \rangle$ to any event $a$ in $p_i$. The entire system of clocks, represented by function $\mathcal{C}$, assigns number $\mathcal{C} \langle b \rangle$ to any event $b$, where $\mathcal{C} \langle b \rangle = \mathcal{C}_j \langle b \rangle$ if $b$ is an event in process $p_j$. Thus, the {\em happens-before} may be restated as follows.

%\begin{itemize}
	%\item 
\vspace{1mm}
\noindent{{\em Clock Condition}}: For events a, b: if $a \rightarrow b$ then $\mathcal{C} \langle a \rangle < \mathcal{C} \langle b \rangle$.
\vspace{1mm}
%\end{itemize}

Here, the converse condition may not hold since that implies any two concurrent events must occur at the same time. Also, according to the {\em happens-before} relation, this Clock Condition is satisfied if the following conditions hold.
\begin{enumerate}[-]
	\item If $a$ and $b$ are events in process $p_i$, and $a$ comes before $b$, then $\mathcal{C}_i \langle a \rangle < \mathcal{C}_i \langle b \rangle$.
	\item If $a$ is a sending of a message by process $p_i$ and $b$ is the receipt of that message by process $p_j$, then $\mathcal{C}_i \langle a \rangle < \mathcal{C}_j \langle b \rangle$.
\end{enumerate}

%\noindent{\bf Vector Clock-Based Causal Ordering.}
\subsection{Vector Clock-Based Causal Ordering}
%talk about how the consistent ordering can be done
%We now discuss how Lamport's logical clock can be used to achieve causal ordering of events in a node during BOID's proof of work protocol execution in that node. 
Since Lamport's logical clock does not guarantee that if $\mathcal{C} \langle a \rangle < \mathcal{C} \langle b \rangle$ then $a \rightarrow b$, it may not be used directly when concurrent events are present. In the following, we discuss our vector clock-based event ordering protocol that leverages Lamport's Logical Clock notion. In each miner, a counter represents a logical clock. Also, each miner $m_i$ maintains a vector $V_i[1 \cdots n]$, where $n$ is the number of miners (including itself) that may trigger an event in it, $V_i[j]$ is the number of messages/actions from miner $m_j$ that has been received at it, and $V_i[i]$ is the number of messages/actions sent by itself. These vectors are similar to the vectors used in the {\em vector clock system}~\cite{vector1, vector2}, however, with the following exceptions. (1) The vector clock system requires each node to have entries for all other nodes in the system, whereas, in T-IoT, a miner maintains entries only for its associated miners. (2) The miners in T-IoT reset their vectors as soon as the associated events are validated. This resetting technique also accounts for the dynamic node join/leave (e.g., installing/uninstalling event chains) in the network. If a node dies during a block validation, the gateway can still stick to the {\em Nakamoto consensus} and decide on approving/disapproving an event. Therefore, if a node leaves/joins, there is no additional overhead.

%Therefore, if a node leaves/joins, there is no additional overhead due to vector resetting

Formally, vector $V_i$ in T-IoT has the following properties.
\begin{enumerate}[-]
	\item All miners initialize their vectors $V[1 \cdots n]$ with 0. 
	\item When miner $m_i$ sends a message, it increments $V_i[i]$ and attaches its vector as a timestamp (TS) to the message.
	\item When miner $m_j$ receives a message from miner $m_i$, $m_j$ sets $V_j[k] = \max(V_j[k], TS_i[k])$, $\forall k \not= j $ and then increments $V_j[j]$  by 1. 
	Here, $TS_i$ is the TS sent by miner $m_i$.
\end{enumerate}

Now, if $a$ is an event from miner $m_i$ and $b$ is an event from miner $m_j$, miner $m_k$ can determine the causal relation between events $a$ and $b$ as $a \rightarrow b$ if $V_k(a)[i] \le V_k(b)[i]$,  where $V_k(a)[i]$ denotes the $i$-th entry of miner $m_k$'s vector after reception of event $a$ from miner $m_i$ and $V_k(b)[i]$ denotes the $i$-th entry of miner $m_k$'s vector after reception of event $b$ from miner $m_j$. However, such causal relation will be true only if the communication channel is deemed reliable and follows the first-in-first-out (FIFO) message forwarding strategy. Thus, the gateway simply follows the FIFO strategy while passing messages between nodes. Miners, however, confirm the causal message delivery as follows.
Miner $m_j$ postpones creation of a block for validation in T-IoT until
\begin{enumerate}[-]
	\item $TS_i[i] = V_j[i] + 1$, where $TS_i$ is the TS sent by miner $m_i$.
	\item $TS_i[k] \le V_j[k], \forall k \not= i$.
\end{enumerate}

As an example, let miner $m_3$ have $V_3 = [0, 1, 1]$, i.e., miner $m_3$ has received 0 message from miner $m_1$, 1 message from miner $m_2$ and, sent 1 message so far. Later, miner $m_1$ sends a message with $TS_1 = [1, 2, 0]$. At this point, miner $m_3$ checks and confirms it is the next message from miner $m_1$ since $TS_1[0] = V_3[0] + 1$. However, miner $m_3$ does not create a block immediately for validation since $TS_1[2] > V_3[2]$. Instead, miner $m_3$ waits for a message from miner $m_2$. Once the missing message is received, miner $m_3$ creates a block for validation with respective order and solves the respective puzzles. As soon as that block is validated, all the associated miners decrease entries in their vectors depending on what events have been validated. Each validated block in this way contains the global causal ordering of the events.
%and the gateway uses this ordering to create the data provenance in T-IoT, as described in the following section.

\section{T-IoT Provenance Creation}\label{sec:provenance}

In this section, we describe how the gateway creates a data provenance in the form of DAGs. This protocol is essential to provide a pervasive knowledge representation of the ordering. 
Additionally, it may serve as the entry point for the developer's community for building applications. %At this time, we design a low-level provenance support and leave the programming platform support as the future work.
To recall, PROV-DM DAGs have {\em entity}, {\em activity}, and {\em agent} nodes. 
%To avoid confusion between these DAG nodes and the IoT nodes (e.g., miners), we refer to the former simply as ``nodes" and the latter as ``miners" in this section.
We refer to these nodes as {\em DAG-nodes}. In parallel to the block validation, the gateway identifies the DAG-nodes continually and generates provenance graphs of the trigger-action-based chains (or simply {\em action-chains}).
An action-chain involves chaining of one or more events in accordance with the {\em happens-before} relation.
In the following, we discuss the components and procedure of the T-IoT provenance protocol.

%\noindent{\bf Device Handlers.}
\subsection{Device Handlers}
They represent the IoT nodes at the gateway. Communication between the gateway and a node (miner/non-miner) happens through the node's device handler. Each device handler manages the low-level commands (e.g., {\em event\_handlers} for the supported events) and exposes a programmable interface that allows the developers to provide custom-built automation support. The {\em event\_handler} for each event is known to the associated nodes and the gateway.
A user may communicate with a device handler to invoke an {\em event\_handler} to execute an action (e.g., lock a door) or subscribe to broadcast events (e.g., motion detection event).

%\noindent{\bf Action-chains.}
\subsection{Action-chains}
The gateway creates a provenance graph for each action-chain.
An action-chain may involve one or multiple events that lead the system to a specific state. 
Each action-chain has a {\em start} event and one or more {\em end} events, where the {\em start} event is the root cause and an {\em end} event is a final outcome. An action-chain that involves only one event will recognize that event as both {\em start} and {\em end} events. Such an action-chain may emerge in the IoT platforms due to the interventions from the users, device malfunction, attackers, and/or relationships derived from all the existing action-chains.

%\noindent{\bf Identifying Entity, Activity, and Agent.}\label{sec:identify_dag_nodes}
\subsection{Identifying Entity, Activity, and Agent}\label{sec:identify_dag_nodes}
%The gateway is responsible for identifying the entities, activities, and agents for provenance creation. 
To recall, an entity is a data object which led the system to its current state, an activity is responsible for creating one or multiple entities, and an agent helps one or multiple activities to create entities. 
In T-IoT, the gateway identifies each event (observable in the system) as an entity, each validated {\em event\_handler} that generated an event as an activity, and each IoT node (e.g., miner) that executes an event as an agent. For example, the sound of an alarm (observable symptom) in a smart home is identified as an event, the invoking function alarm\_on is identified as an activity, and the alarm sensor is identified as an agent. 
To define the dependencies between the DAG nodes, the gateway encodes appropriate edge labels, as depicted in Figure~\ref{fig:prov}.

%\noindent{\bf Provenance Creation.}
\subsection{Provenance Creation Algorithm}
Based on the observable symptoms, the gateway first identifies the {\em end} event/s. 
Note that it can start with any number of such independent events which eventually conform to a single or multiple action-chains. 
For a single event, the gateway locates it in the Blockchain and identifies the corresponding entity, activity, and agent. A recent event may reside within a recent block. The gateway thus starts from the latest blocks of each chain of blocks in the T-IoT Blockchain. The dependencies between the IoT nodes are known to it. When the {\em end} event/s are found, it follows the events in a block in the reverse order and looks for a triggering event. If it needs to traverse multiple blocks, it does so in the reverse order as well through the selected chain of blocks. 
%This {\em backtracking}-based search procedure ends when all the observable symptoms (e.g., events) merge to a single triggering event, finish independently by identifying their own triggering events, or/and both.
This procedure ends when all the observable symptoms merge to a single triggering event  or finish independently by identifying their own triggering events.
The gateway thus have the provenance graphs of one or multiple action-chains.

\section{Implementation}\label{sec:imple}
\begin{figure}[!htbp]
\centering %\vspace{-0.1in}
\includegraphics[width=0.4\textwidth]{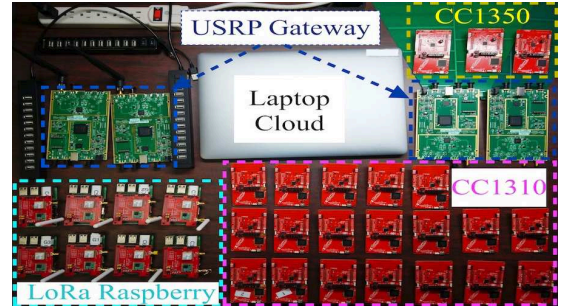} %\vspace{-0.1in}
\caption{Devices used in our implementation.} %\vspace{-0.1in}
\label{fig:implementation}
\end{figure}

We have implemented a proof of concept IoT platform using GNU Radio~\cite{gnuradio}, USRP~\cite{usrp}, laptop, and several COTS IoT devices including TI CC1310, TI CC1350, and Dragino LoRa Hat on Raspberry Pi 3. GNU Radio is a signal processing toolkit (installed on a PC) for implementing software-defined radios, which is used in our gateway. Our gateway is equipped with four different types of wireless communication technologies: Wi-Fi, BLE, LoRa, and IEEE 802.15.4g. Each technology is supported by a half-duplex USRP device to act as the radio front-end. 
We thus have a multi-radio gateway similar to many commercially available smart home gateways~\cite{logging1}.
We use a laptop that acts as the Cloud backend. The gateway connects to the Cloud backend via Wi-Fi~\cite{gwifi}. 
We have used 19 TI CC1310 devices, 3 TI CC1350 devices, and 8 Dragino LoRa Hat devices as the IoT nodes in our platform. Figure~\ref{fig:implementation} shows the actual devices (except the PC running GNU Radio) used in our implementation.

Each TI CC1310 device is connected to the gateway via IEEE 802.15.4g and uses a CSMA/CA (carrier sense multiple access/collision avoidance)-based MAC (media access control) protocol~\cite{gzigbee}. Each TI CC1350 device is connected to the gateway via BLE and uses the GATT (Generic Attribute Profile) data transfer protocol~\cite{gble, gble2, gatt}. Each LoRa Hat device is connected to the gateway via LoRa communication technology and uses a pure ALOHA-based MAC protocol~\cite{glora}.
We have implemented the T-IoT Blockchain and ordering protocols in each of the IoT devices. 

\section{Evaluation}\label{sec:experiments}
%In this section, we evaluate the performance of T-IoT protocols using our implementation (Section~\ref{sec:imple}). 
In the section, we evaluate all the T-IoT protocols using our implementation described in Section~\ref{sec:imple}.

\begin{figure*}[!htbp]
    \centering %\vspace{-0.1in}
      \subfigure[Storage growth\label{fig:block_storage}]{
    \includegraphics[width=0.28\textwidth]{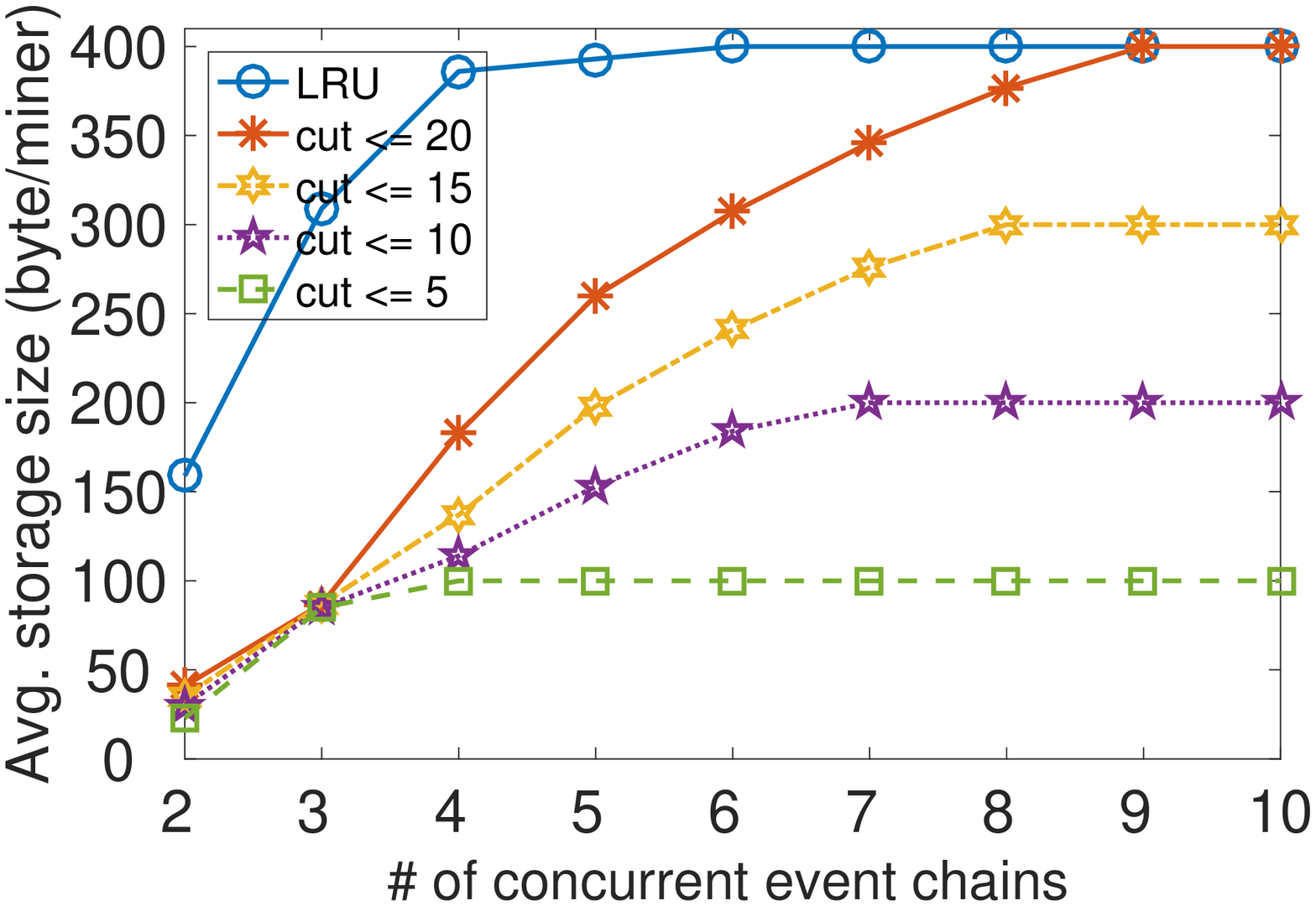}
      }\hfill
      \subfigure[Block validation latency \label{fig:block_latency}]{
        \includegraphics[width=.28\textwidth]{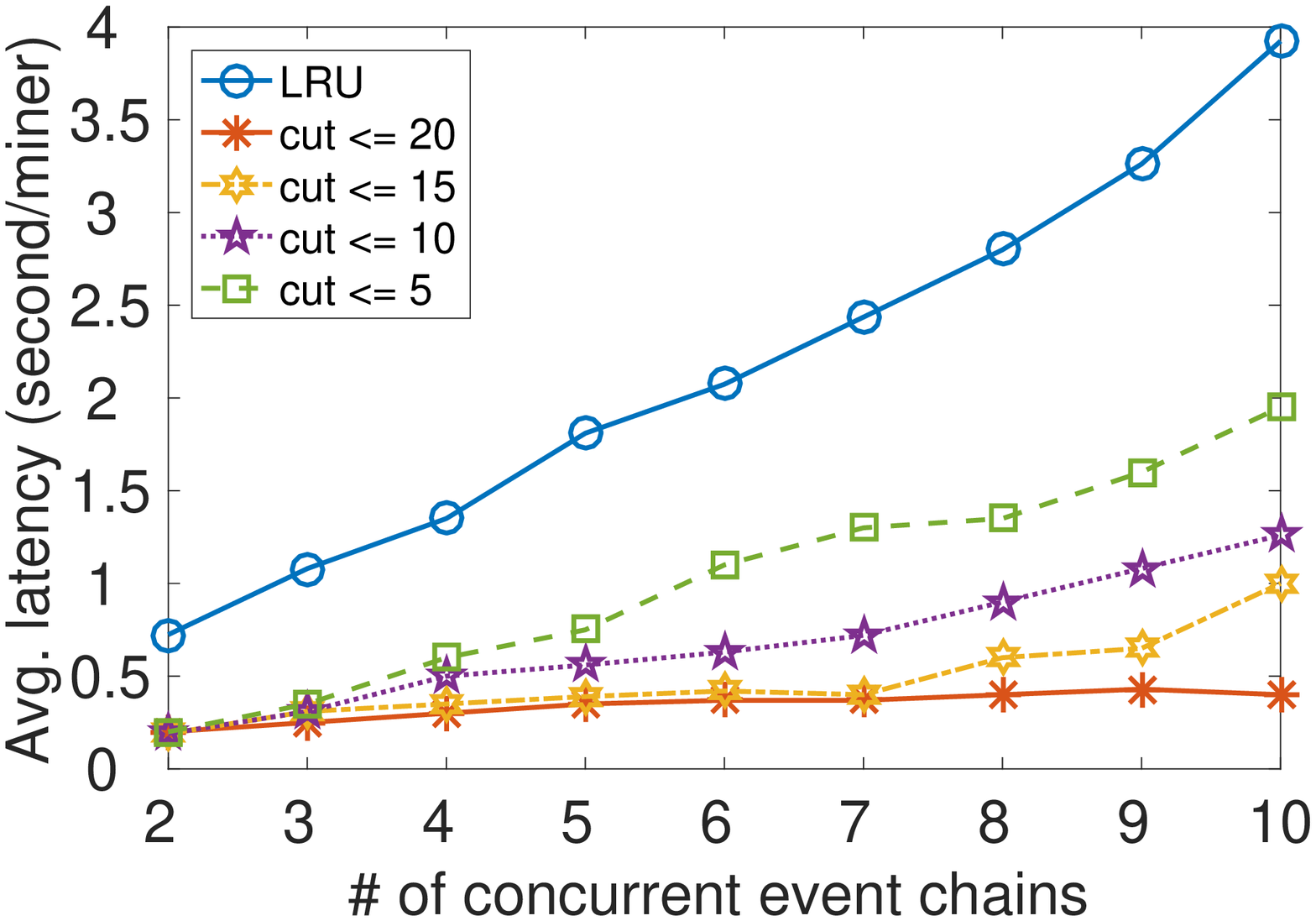}
      }\hfill
      \subfigure[Block validation energy consump.\label{fig:block_energy}]{
        \includegraphics[width=0.28\textwidth]{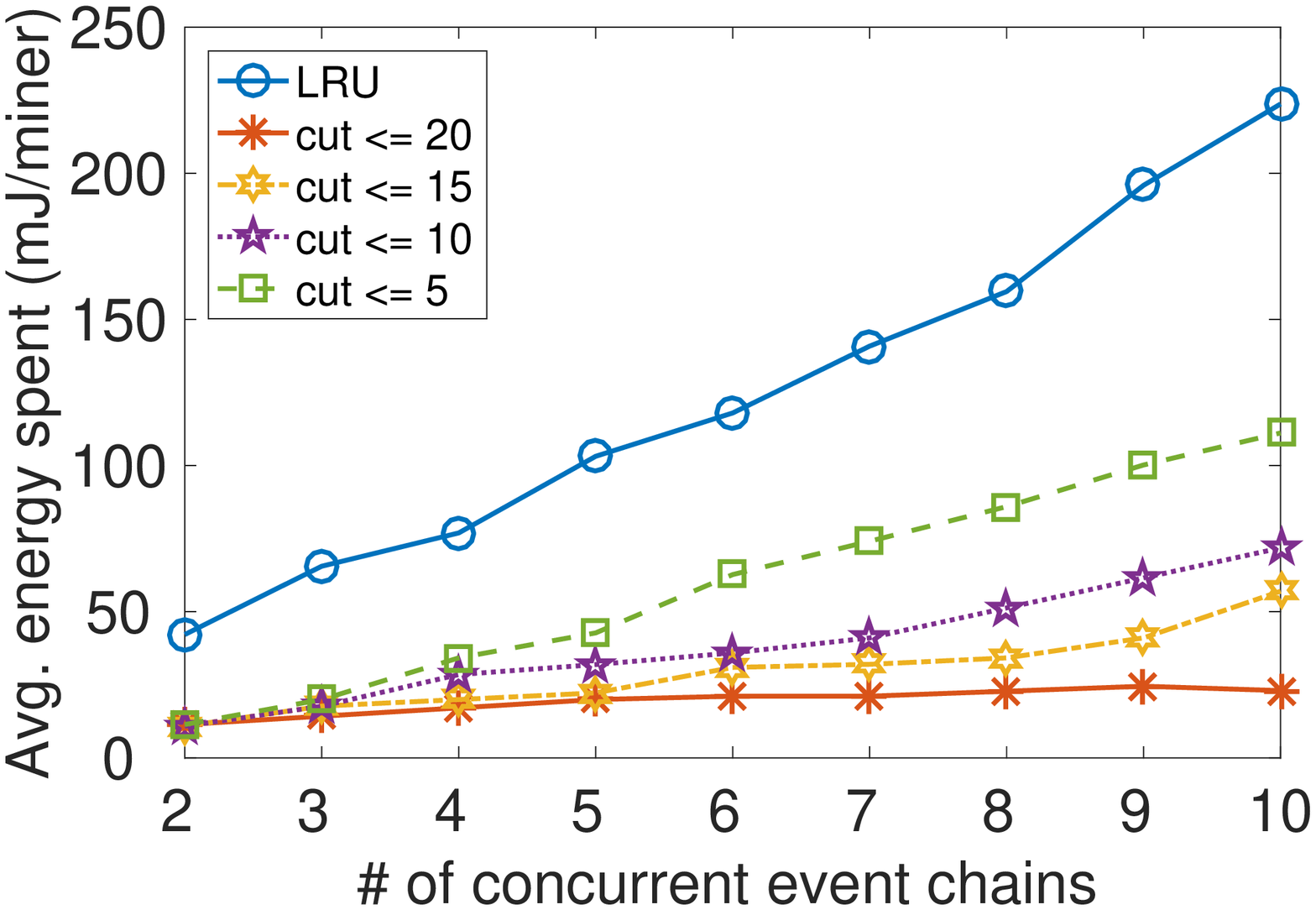}
      } %\vspace{-0.1in}
    \caption{Performance of T-IoT Blockchain protocol with various numbers of concurrent trigger-action-based event chains.} %\vspace{-0.15in}
    \label{fig:block}
\end{figure*}
\subsection{Experimental Setup}\label{sec:setupchain}

\noindent{\bf Event Chains.}
We let each IoT node execute one unique event (total of 30 events, which is typical in the smart homes~\cite{lhome, dataset}).
%We thus have 30 events in our setup which is similar to a typical large smart home scenario~\cite{lhome, dataset}. 
Each event has a unique {\em event\_handler} (i.e., number). 
We also pseudo-randomly create 10 trigger-action-based event chains of lengths 10--20 (thus, the maximum size of a partial cut will be 20), where a device may participate in 5 chains at maximum (also typical in the smart homes). In each chain, we find the first event and (re)activate it during our experiments (reactivation interval: 10--20 seconds). 
Note that we are unable to find datasets, generated by the existing IoT platforms, which fit and account for the novelty of our design. 
%In the experiments, we thus crafted the above event chains involving all the devices in our testbed.

\noindent{\bf Device Attributes.}
%Typically, a sensor node uses a very small payload to communicate with the gateway. For example, a ZigBee temperature node may only use $\approx 30$ bytes in its payload~\cite{snow_cots}. 
Each node uses a 30-byte payload (typical for sensors~\cite{snow_cots}) while the 
actual frame size may vary depending on its communication protocol. The channel bandwidths for BLE, IEEE 802.15.4g, and LoRa (spreading factor: 9, coding rate: $\frac{4}{5}$) are 1 MHz, 98 KHz, and 500 KHz, respectively. These settings let each technology take approximately the same time to send a 30-byte payload. Each node also uses a 15 dBm of transmission power.
For storage capacity, we find that $\approx$ 2560 bytes of the TI CC13x0 (CC1310 or CC1350) are usable, which again shrinks as the main-thread-stack grows (up to 1024 bytes). Considering the program size, we thus limit a maximum of 20 blocks to be saved in the node's flash memory, which is 400 bytes (explained in the next paragraph). 
The gateway is connected to a laptop (via Wi-Fi) where it saves the entire ledger.
%The Wi-Fi data rate is very high and the associated communication latency is negligible compared to the IoT nodes. 
It also chooses 30 random primes between the 800th and 1200th primes, calculates their primitive roots, and assigns random roots to the IoT nodes, as required.

\noindent{\bf Blockchain Parameters.}
Typically, the maximum size of a block and the maximum number of transactions per block are fixed (e.g., 1-MB block and 400 transactions per block in Bitcoin~\cite{nakamoto2008bitcoin}).
%For example, in Bitcoin ecosystem, a miner can generate a block of maximum 4000 transactions, while not exceeding the block size of 1 MB~\cite{nakamoto2008bitcoin}. 
In T-IoT, we limit the size of a block to 20 bytes to fit inside the payload of a message. The rest of the 10 bytes are used to encode $K$ of our PoW puzzle. In our setup, a device may be added to at most 5 event chains, and thus may try to validate 5 events at maximum in one block. We reserve 2 bytes for encoding the value of $K$ so that a node can fit five $K$s, along with a block inside a payload. With 2 bytes, the value of $K$ ranges between 0 and $2^{16}$ (unsigned). The above limits may provide reasonable protection against the compromised nodes and can be changed if needed. 
For the block size, we allow a maximum of 20 bytes, where the block ID$<i,j>$ is 16 bits (8 bits for each index), block hash is 8 bits, parent hash is 8 bits, transaction hash is 8 bits, and each transaction is 8 bits. Without the transactions, the size of a block adds up to $(16 + 8 + 8 + 8) = 40$ bits. Thus, leaving the space for a maximum of  $(20 * 8 - 40)/8 = 15$ transactions (each of 8 bits) in a block. With these setup, T-IoT can host $2^{8}$ distinct events in any IoT platform.
The hash values in the experiments are calculated based on the XOR function.

\begin{figure*}[!htbp]
    \centering %\vspace{-0.1in}
      \subfigure[Message-to-event ratio\label{fig:order_message}]{
    \includegraphics[width=0.28\textwidth]{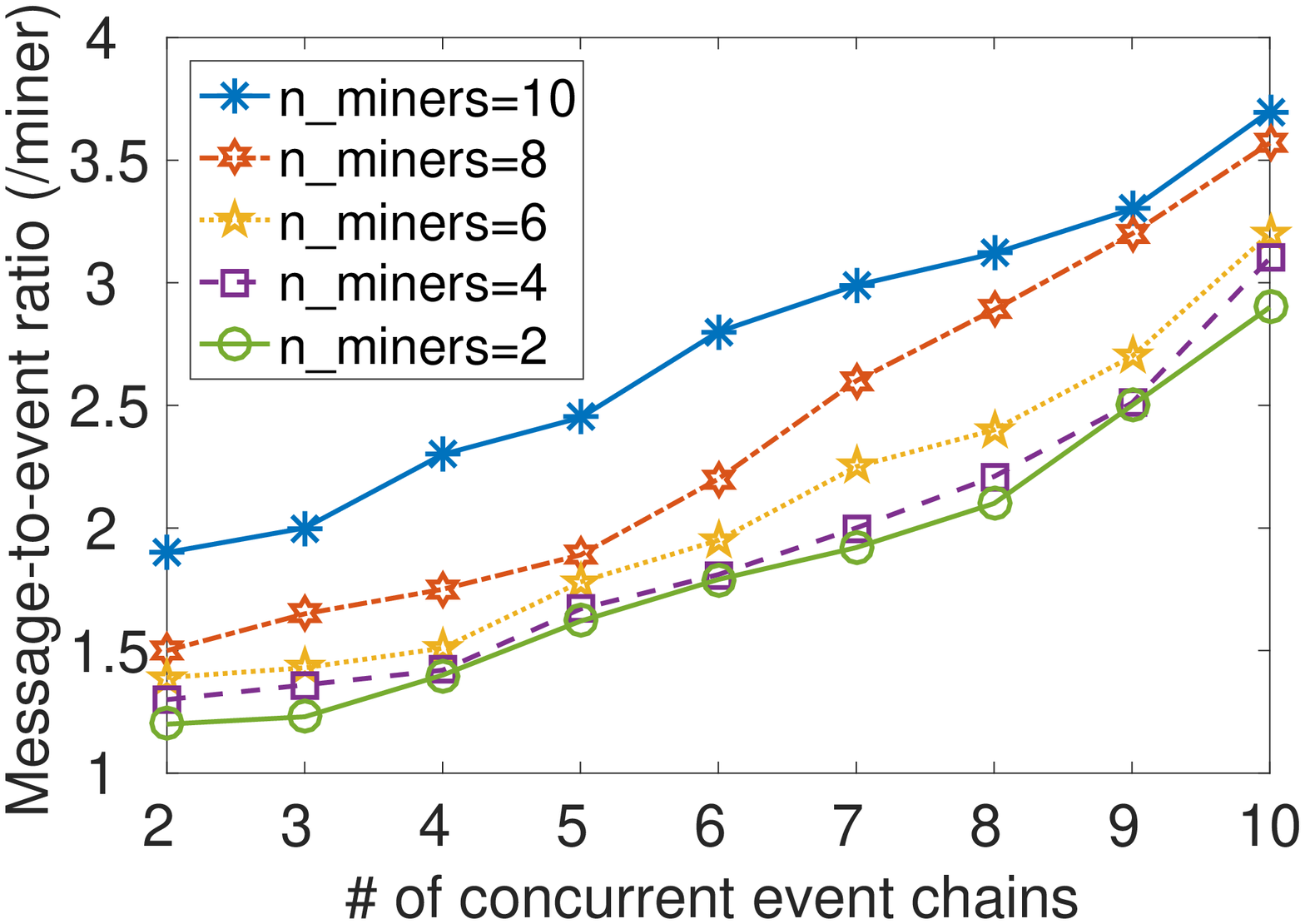}
      }\hfill
      \subfigure[Event ordering latency\label{fig:order_latency}]{
        \includegraphics[width=.28\textwidth]{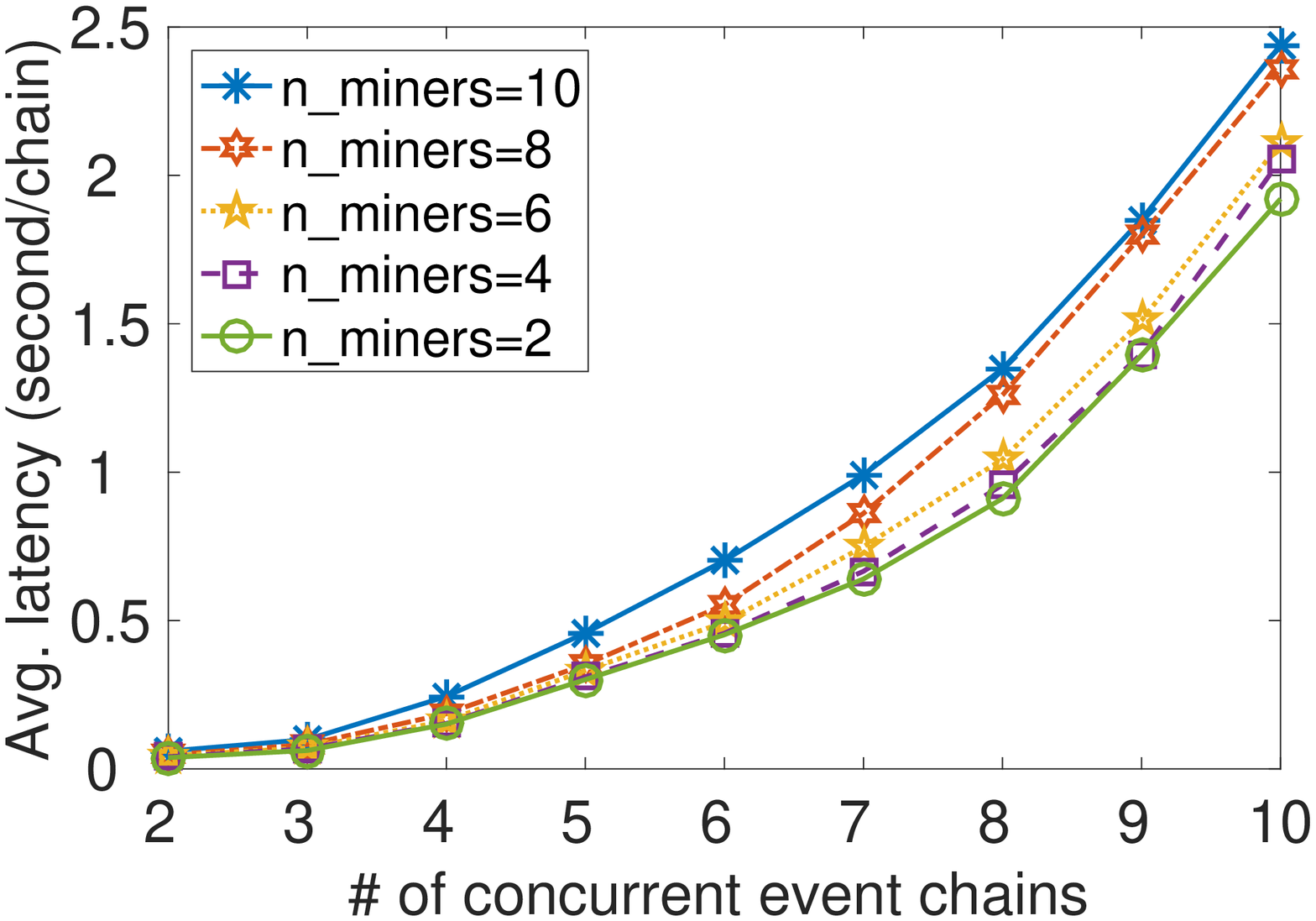}
      }\hfill
      \subfigure[Event ordering energy consump.\label{fig:order_energy}]{
        \includegraphics[width=0.28\textwidth]{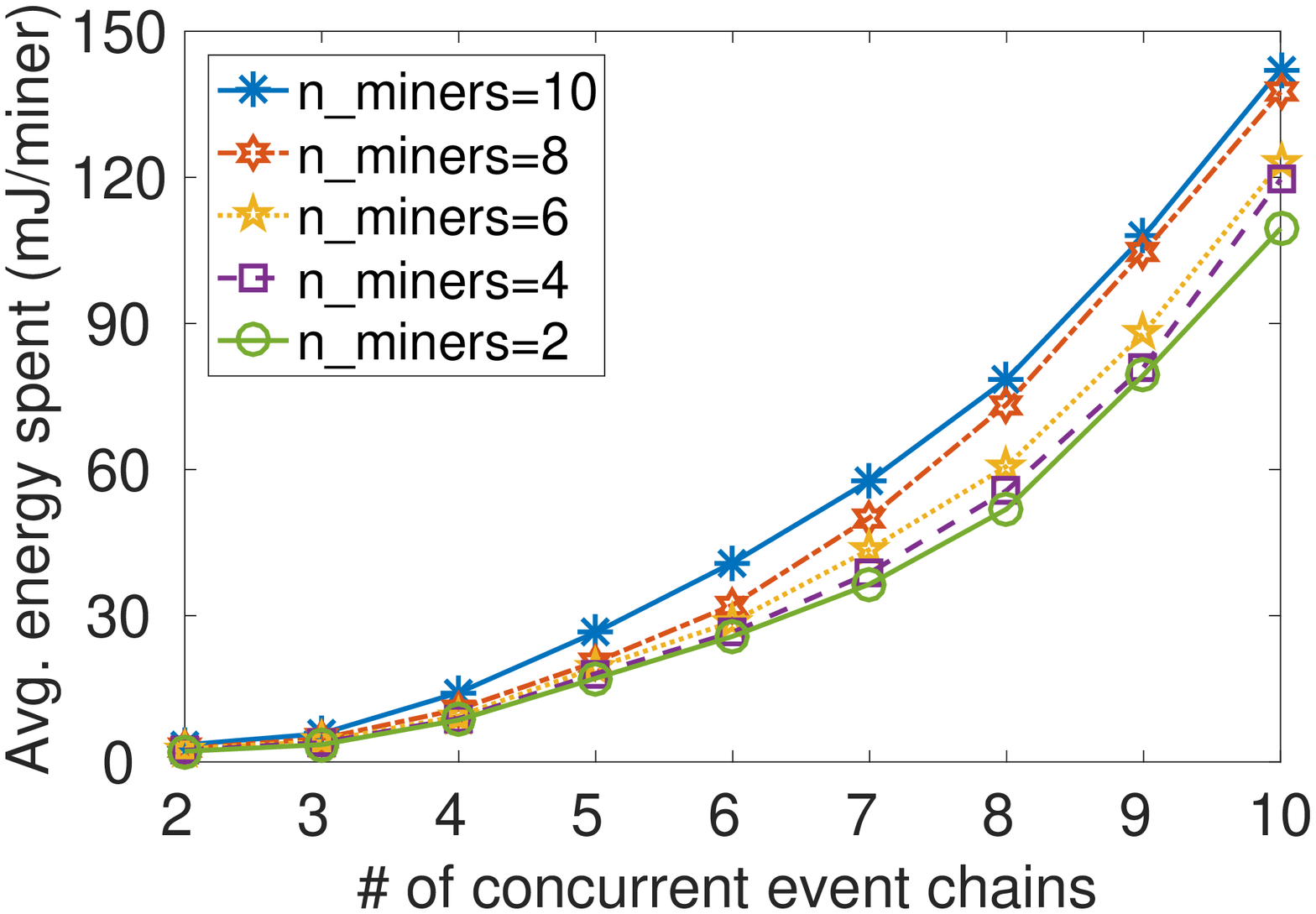}
      } %\vspace{-0.1in}
    \caption{Performance of the T-IoT event ordering protocol under various numbers of concurrent trigger-action-based event chains.} %\vspace{-0.25in}
    \label{fig:order}
\end{figure*}
\subsection{Performance of the T-IoT Blockchain} \label{sec:bchainper}
We now evaluate the performance of the T-IoT Blockchain protocol in terms of the storage growth, latency, and energy requirements in the miners. We allow between 2 to 10 different chains to execute in parallel. Each node associated with an executing chain acts as a miner. Miners, however, are allowed to save and replace block/s that have events belonging to a partial consistent cut of size less than or equal to a fixed number C. We repeat this experiment 5 times by setting the value of C to 1, 5, 10, 15, and 20, respectively. Setting C = 1 refers to the LRU replacement policy, which is the {\em baseline} for comparison since it is the naive approach.

\noindent{\bf Storage Growth.}
As shown in Figure~\ref{fig:block_storage}, the average storage size per miner is approximately 160 bytes when 2 chains execute concurrently and the miners store/replace blocks based on the LRU policy, compared to 23, 30, 35, and 41 bytes when they store/replace blocks based on the cut sizes $\le$ 5, 10, 15, and 20, respectively. As the number of concurrent chains increases, the average storage size per miner also increases for all the scenarios. LRU, however, saturates the miner's storage capacity faster (with 4 concurrent event chains) than the other scenarios. Between other scenarios, each cut gradually approaches its limit, e.g., the storage in a miner having a cut size $\le$ 5 does not grow beyond 100 bytes ($5 *$ 20 bytes). This experiment thus shows that the LRU replacement policy performs the worst. Additionally, with a cut size $\le$ 5, the miners can easily execute the T-IoT blockchain protocol.

\noindent{\bf Validation Latency.}
Figure~\ref{fig:block_latency} shows the average latency per miner per block. When 2 concurrent chains execute, the average latency per miner is approximately 0.72 seconds when the miners save/replace blocks using the LRU policy, compared to 0.2, 0.19, 0.19, and 0.2 seconds when they use the cut sizes $\le$ 5, 10, 15, and 20, respectively. As the number of concurrent event increases, the latency increases for all the cases. Again, LRU policy performs the worst (e.g., the latency is approximately 4 seconds for 10 chains). The average latency per miner with a cut size $\le$ 20 is approximately 0.4 seconds (for 10 chains), which is very low compared to the others. In fact, the change in latency is negligible across different concurrent chains. This experiment thus confirms that miners may validate blocks faster if the consistent cut size is larger.

\noindent{\bf Validation Energy Consumption.}
As shown in  Figure~\ref{fig:block_energy}, when 2 concurrent chains execute, the average energy consumption per miner to validate one block is approximately 42.2 mJ for the LRU policy, compared to approximately 11.4, 10.83, 10.9, and 11.4
mJ for the cut sizes $\le$ 5, 10, 15, and 20, respectively. The increase in energy consumption per miner per block follows the similar trend of the average latency, as the number of concurrent chains increases. Overall, for 10 chains, a miner with cut size $\le$ 20 consumes the minimum energy (approximately 23 mJ) compared to others since it needs no block replacement. Hence, an increase in the cut size increases the energy efficiency of the T-IoT Blockchain protocol.

\subsection{Performance of the Ordering Protocol}
In this section, we evaluate the performance of the ordering protocol in terms of {\em message-to-event ratio} (MER), latency, and energy requirements in the miners. MER is defined as the ratio of the number of messages to the number of events in a chain.
%Our setup (e.g., event chains) for this experiment is similar to those in Section~\ref{sec:bchainper} with the following exceptions. 
In experiments, we set the consistent cut size to $\le$ 20 in all the miners since it is the most energy-efficient and requires the least time to validate a block. Also, we vary the number of miners between 2 and 10 for each event chain to determine its effects on the ordering.

\noindent{\bf Message-to-event Ratio.}
As shown in Figure~\ref{fig:order_message}, when 2 concurrent chains execute, the MER per miner is approximately 1.2, 1.3, 1.39, 1.5, and 1.9 for 2, 4, 6, 8, and 10 miners, respectively. As the number of chains increases, the MER per miner also increases almost linearly for all the cases. Also, as we increase the number of miners, the MER per miner increases, which is due to an increase in the number of messages between the miners. For example, in the case of 10 miners and 10 chains, the MER per miner is approximately 3.7, which is practical with respect to its latency and energy requirements, as discussed below.

\noindent{\bf Event Ordering Latency and Energy Consumption.}
As the MER increases with the number of miners, the event ordering latency also increases. Figure~\ref{fig:order_latency} shows that when 10 concurrent chains execute,  the average latency per chain (i.e., average latency per miner) is approximately 1.92, 2.05, 2.11, 2.36, and 2.44 seconds for 2, 4, 6, 8, and 10 miners, respectively. Such a sub real-time latency is practical for the smart home, traffic, or/and crime monitoring applications.
As shown in Figure~\ref{fig:order_energy}, the average energy consumption per miner follows the similar trend that we observe in the average latency. For 10 chains, the average energy consumption per miner is approximately 109, 120, 123, 137, and 142 mJ as we set the number of miners to 2, 4, 6,  8, and 10, respectively, which is also practical for battery-powered IoT nodes. 
%In general, it is better to have less number of miners in T-IoT, as per Figure~\ref{fig:order_energy}.

\subsection{Experiments on Provenance Creation}
In this section, we experiment on the T-IoT provenance creation protocol that runs at the gateway. Specifically, we show that the provenance creation protocol is accurate and timely. Additionally, we represent the achieved ordering in the form of PROV-DM DAG, which may be perceived as an observable output of the T-IoT framework. In the following, we first discuss the setup and then describe the experimental results with a provenance graph.

\noindent{\bf Setup.} To facilitate this experiment, we take out 5 of our devices from the experimental setup of the event chains (Section~\ref{sec:setupchain}) and label them as different sensor nodes such as smoke detector, smoke monitor, window sensor, fire alarm sensor, and water sprinkler sensor. These sensors have the following relationships. The smoke monitor sensor activates the window sensor, fire alarm sensor, and water sprinkler  as it observes the smoke detector detects smoke. The above setup is thus an event chain that associates 5 sensors. Note that the smoke monitor does not execute any event by itself, but activates events in the window, alarm, and sprinkler by invoking their {\em event\_handlers}. Additionally, these are all fabricated events (thus, no safety hazards) in our setup where the smoke detector generates a smoke detection event as soon as we start our experiment. To show the correctness of the T-IoT provenance creation protocol, we run our smoke event chain in parallel to the 10 other concurrently executing event chains where the cut size is set to $\le$ 20 and the number of miners is set to 5. Note that the T-IoT Blockchain and event ordering protocols also run in the background. After 15 seconds from the start of our experiment, we execute a user program at the gateway that asks for a provenance graph of the window open, alarm on, and sprinkler on events. We repeat the same experiment 10 times with a random interval between 10 to 15 seconds to show its scalability and correctness.

\begin{figure}[!htbp]
\centering %\vspace{-0.15in}
\includegraphics[width=0.49\textwidth]{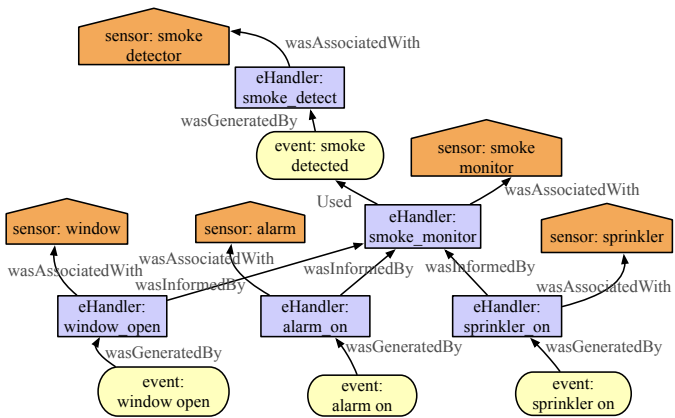}%\vspace{-0.1in}
\caption{Visual representation of the smoke detect provenance graph.} %\vspace{-0.12in}
\label{fig:smoke} %\vspace{-0.15in}
\end{figure}

\noindent{\bf Results.}
The gateway creates and returns the provenance graph in the form of a DAG for each run of our experiment with an accuracy of 100\% and average latency of approximately 3.5 seconds. We draw the provenance graph in the form of PROV-DM provenance model which is shown in Figure~\ref{fig:smoke}. As shown in this figure, all of our three  requesting events, i.e., window open, alarm on, and sprinkler on converge to their root cause, which is a smoke detection event. This experiment thus shows that the provenance creation protocol in T-IoT is pervasive, accurate, and timely. 
%This experiment thus shows that the provenance creation protocol in T-IoT is pervasive, accurate, and timeliness.

\section{Related Work}\label{sec:related}

\noindent{\bf Blockchain in IoT.}
Blockchain protocols have been adopted in the IoT platforms in various ways that include secure data transfer between the gateway and Cloud~\cite{lin2017using},  IoT device management~\cite{huh2017managing, danzi2018analysis, novo2018blockchain, ourad2018using}, securing multiple smart homes collectively~\cite{dorri2017towards}, secure data sharing within/across organizations~\cite{hashemi2016world, zyskind2015decentralizing}, and proposing business models for IoT~\cite{zhang2017iot}. This large body of works, however, adopts the Blockchain protocol on the overlay (or overhaul) network where the devices (e.g., gateway, router, servers, etc.) have sufficient computing power and storage capacity and are connected via fast Internet connection, thus cope with the Blockchain requirements (e.g., SpeedyChain~\cite{kang2019toward} and EdgeChain~\cite{pan2018edgechain}).

In this work, we bring the Blockchain protocol to the end devices (e.g., IoT sensor/actuator nodes deployed for sensing/actuation) of the IoT platforms. To the best of our knowledge, ours is the first Blockchain protocol that is tailored for the resource-constrained (e.g., limited storage and computational power) heterogeneous IoT devices  that connect to the IoT gateway via wireless.
Additionally, to the best of our knowledge, T-IoT is the first framework that interconnects the Blockchain and vector clock to provide {\em transparent and tamper-proof} event ordering in general. 
A few works that bring distributed ledger or Blockchain to the IoT devices include IOTA~\cite{iota} and Sensor-Chain~\cite{shahid2019sensor}. IOTA is a cryptocurrency for the IoT industry, which maintains a distributed ledger (not Blockchain) among homogeneous (e.g., communication protocol) devices with fixed types of preloaded transactions. Sensor-Chain Blockchain does not incorporate any PoW consensus protocol, works within homogeneous nodes, and cannot guarantee tamper-proof ledger maintenance. In contrast to these, T-IoT enables Blockchain over heterogeneous IoT nodes to provide tamper-proof transparency in the IoT platforms.

\noindent{\bf Data Provenance in IoT.}
Data provenance in the IoT platforms can be broadly categorized into {\em device-centric}~\cite{aman2017secure} and {\em platform-centric} \cite{logging1} models. Device-centric provenance reflects causal relationships of data objects within a device that cannot be generalized for building a global provenance of its embodying platform due to the heterogeneity of the devices. In this paper, we facilitate platform-centric provenance that supports several degrees of heterogeneity.
%including manufacturer and communication protocol.

The closest work with the similar goal to ours is~\cite{logging1} that also provides a platform-centric provenance for the IoT platforms. However, we have the following differences.
(1) The framework in~\cite{logging1} is orchestrated by instrumenting/adding software programs at different levels such as the platform Cloud/gateway and the user applications (similar to those in \code{ifttt}~\cite{ifttt} and \code{tray.io}~\cite{tray}). On the other hand, T-IoT is designed by instrumenting the device's programmable interfaces (e.g., device handlers).
Additionally, our device instrumentation leaves the doors open for building innovative protocols for the IoT platforms (e.g., Blockchain and vector clock) that are not possible in~\cite{logging1}. 
(2) We enable Blockchain in the resource-constrained heterogeneous IoT devices that is unique in T-IoT. (3) For ordering of events, we customize the vector clock system while~\cite{logging1} depends on instrumenting the application programs. Additionally, our event ordering logically synchronizes the IoT devices, which is not possible in~\cite{logging1}. 
(4) For any change in the existing set of trigger-action-based event chains,~\cite{logging1} will need to re-instrument program code, which is unrealistic. On the other hand, in T-IoT, it can be handled effectively by enabling/disabling {\em event\_handlers} from device handlers. Thus, T-IoT is more scalable.

\section{Conclusions}\label{sec:conclusion}
In this paper, we have proposed a transparent and tamper-proof event ordering framework called T-IoT by tailoring the Blockchain protocol for the resource-constrained IoT devices.
To overcome their storage and computation limitations, we have allowed the devices to save only a portion of the ledger based on a partial consistent cut and engineered an efficient modular arithmetic-based PoW puzzle, respectively. 
Ordering of the events has been achieved through the adoption of the vector clock system, customized for the IoT platforms. We have then proposed a backtracking-based data provenance creation protocol. We have implemented T-IoT using COTS devices. Our experiments with 10 concurrent trigger-action-based event chains (each chain involving up to 20 devices and each device participating in 5 different event chains) have demonstrated that the ordering of these events may be done in 2.5 seconds at the cost of 140 mJ of energy per device, which is much promising for many IoT applications, including smart home, traffic/accident monitoring, and crime investigation. 

\section*{Acknowledgments}
This work was supported through NSF grants CAREER-2211523 and CNS-2211510.

\bibliographystyle{IEEEtran}
%\balance
\bibliography{IEEEabrv,blockchain-short}

\vfill

\end{document}